\documentclass[12pt]{article}
\usepackage[margin=0.8in]{geometry}
\usepackage{authblk}
\begin{document}

\title{A Shiny Application for Conducting Electronic Surveys Using Randomized Response Techniques}
\author[1]{G.N. Singh} 
\author[2$^*$]{D. Bhattacharyya} 
\author[3]{A. Bandyopadhyay}
\affil[1]{\footnotesize Department of Mathematics \& Computing,
	Indian Institute of Technology (Indian School of Mines),
	Dhanbad-826 004, Jharkhand, India. Email: gnsingh\_ism@yahoo.com, gnsingh@iitism.ac.in }
\affil[2]{\footnotesize Department of Mathematics \& Computing,
	Indian Institute of Technology (Indian School of Mines),
	Dhanbad-826 004, Jharkhand, India. Email: diya\_bhattacharyya@yahoo.co.in 
}  
\affil[3]{\footnotesize Department of Mathematics, Asansol Engineering College, Asansol-713305, India. \newline  Email: arnabbandyopadhyay4@gmail.com}
\date{}

\maketitle

\begin{abstract}
Randomized response techniques (RRT) are useful for collecting information on sensitive or confidential attributes in sample surveys. However, such RRTs are rarely used except for pure academic research, as they are deemed to be confusing for respondents and difficult to be administered. Switching to an electronic medium is a solution for this. Standard electronic survey mediums lack the option of effectively incorporating a randomization device into it. This paper introduces a Shiny application that implements two popular Randomized Response Techniques, namely, Warner (1965) and Simmons (1967), to collect information on sensitive issues via electronic surveys, and save them using local or remote storage. Code and its detailed explanation has been provided for suitable use by survey statisticians and practitioners. The method suggested here can be extended to implement any Randomized Response Technique. Thus, this paper lays the groundwork of implementing RRTs as electronic surveys for widespread use by the polling industry.
\end{abstract}

\textbf{Keywords:} Randomized response technique, sample surveys, electronic survey, shiny, R.

\section{Introduction}


Sensitive or confidential questions are often a part of sample surveys. Unlike innocuous questions, which generally receive proper responses, interviewees are often reluctant to disclose information related to stigmatized matters. This is specially true for surveys related to socially relevant issues, such as whether a high school student smokes, or whether an unmarried woman had an abortion, etc. It is also true for the recent Covid-19 pandemic, as individuals often hide their symptoms for fear of social isolation, discrimination, etc. (\cite{dishonesty})\\

One way to overcome hesitance from the respondent is to design methods that protect his privacy and reveal no identifying information to the interviewer. This is achieved by making use of randomization devices, such as dice, coins, spinners, deck of cards, etc.\\

Surveys were traditionally conducted via face-to-face interviews, mails, or telephonic conversations. However, there has been a recent shift towards electronic surveys due to increased response rates, lower costs, and administrative convenience. Such medium is specially suitable for surveys on sensitive issues, as it provides confidence of anonymity in absence of an interviewer. Electronic surveys are notably useful in times of global pandemics such as Covid-19, when human contact should be minimized as far as possible, and people are particularly cautious about exposure to interviewers or questionnaires sent via mail.\\

The chief merit of conducting surveys employing RRT via an electronic medium, however, is in its simplicity and effectiveness. As noted by (\cite{blair}), there are astonishingly few applications of RRTs, despite the advances in the field over a few decades. Use of RRTs has been chiefly limited to pure academic research. This may be attributed to the difficulty of implementing the various methods and the confusion it may cause for the respondents. Often, RRTs are elaborate, multi-stage, with complicated instructions for the respondents. Switching to an electronic medium and simulating the randomization device removes all the complications of the method, while retaining all its advantages. The working of the devices now happen "behind the scenes", and the interviewees are only presented with the final statement obtained from the device, to which they have to answer with a simple "Yes" or "No". The entire process now becomes simple and hassle-free. \\

A number of tools and websites are available to the survey statisticians and practitioners for conducting sample surveys online. However, these surveys are merely capable of providing questionnaires to interviewees and collecting information. There is no scope for incorporating randomization devices, which are essential for conducting surveys involving sensitive attributes.\\

This manuscript proposes tools to conduct surveys involving Randomized Response Techniques (RRT) using Shiny-based applications. The authors have chosen to develop the tools as a Shiny application, which is web-based, instead of an add-on R package for their easy implementation to conduct web-based surveys. The novelty of the manuscript lies in the fact that such tools are not yet available for the survey statisticians, to the best of the authors' knowledge, and the implementation of RRT as electronic surveys has the potential to revolutionize how surveys are conducted.\\

The methods for implementing Warner (1965) (\cite{warner}) and Simmons (1967) (\cite{simmons}) RRTs as web-based survey have been discussed in details in the manuscript. The authors chose to demonstrate the development of web-based surveys using these two particular RRTs because they are widely known, uncomplicated in nature and easy to understand for even the less experienced. However, the application developed here can be easily modified to implement any Randomized Response Technique and conduct surveys on any sensitive attribute. Hence, it has wide-spread applications for socio-economic, political, and other surveys.\\

The manuscript has been divided into the following sections: Section \ref{rrt} briefly recollects the Warner (1965) \cite{warner} and Simmons (1967) \cite{simmons} Randomized Response Techniques, with the corresponding R code for simulating the randomization devices. Section \ref{storage} briefly discusses the merits and demerits of local and remote storage. Code for an application involving local storage has been discussed in details in Section \ref{local}. Remote storage option has been explored in Section \ref{gdrive}. Recommendations for survey statisticians have been put forward in Section \ref{discussions}. \\

The applications developed here have been built in R (\cite{R}), using the packages Shiny (\cite{shiny}), and googlesheets4 (\cite{googlesheets}). The authors have used R version 3.6.2 (2019-12-12) -- "Dark and Stormy Night" for the purpose. \\

\section{Simulation of the Randomization Device:}
\label{rrt}

\subsection{A revisit of Warner (1965) and Simmons (1967) RRT:}

\cite{warner} pioneered the randomized response technique for estimating the proportion of individuals in a population who possess a sensitive attribute. His randomization device contains two statements- "I possess sensitive character S", and "I do not possess sensitive character S". Either statement is displayed, depending on the outcome of the randomization device. For example, if a Coin is tossed and lands on "Head", the first statement may be displayed, while "Tails" may correspond to the second statement. The interviewee only has to answer
"Yes" or "No" to the interviewer, without revealing his actual status with regard to the sensitive attribute.\\ 

\cite{simmons} extended\cite{warner} model by introducing the use of an 
unrelated question about a non-sensitive characteristic. Here, the statements used are "I possess sensitive character S" and "I possess non-sensitive character Y".\\

\cite{warner} RRT has the restriction that the parameter of the randomization device $p \neq \frac12$. \cite{simmons} has no such restrictions.\\

\subsection{Functions for Randomization Devices:}

The Randomization Devices used by Warner (1965) \cite{warner} and Simmons (1967) \cite{simmons} can be easily simulated in R with the help of the function \texttt{sample()}. The survey statistician has complete control over the parameters for the randomization devices, namely, the probabilities of the randomization devices displaying the statement bearing the sensitive attribute. For a single randomization device, this parameter is named "p", while it is named "p1" and "p2" for two randomization devices.\\

The function that simulates the randomization devices is illustrated with the help of a simple example. Suppose it is of interest to discretely estimate the proportion of individuals in a population who have high fever (sensitive attribute), a symptom of Covid-19. The function that simulates the randomization device proposed by Warner (\cite{warner}) is defined as:

\begin{verbatim}
randrt<-function()
{
r<-sample(1:2,1,replace=F,prob=c(p,1-p))
if (r==1)
txt<-"I have high fever." #Set your sensitive attribute here
else if (r==2)
txt<-"I do not have high fever." #Set your non-sensitive attribute here
return(txt)
}
\end{verbatim}

Here, "r" presents the outcome of the Randomization device, and accordingly, the relevant statement to be displayed is assigned.\\ 

Similarly, the function that simulates the randomization device for the unrelated question model proposed by Simmons (\cite{simmons}) is defined as

\begin{verbatim}
randrt<-function()
{
r<-sample(1:2,1,replace=F,prob=c(p,1-p))
if (r==1)
txt<-"I have high fever." #Set your sensitive attribute here
else if (r==2)
txt<-"I do not have high fever." #Set your non-sensitive attribute here
return(txt)
}
\end{verbatim}

When the proportion of individuals possessing the non-sensitive attribute Y is unknown, two randomization devices are utilized. Hence, two functions, randrt1() and randrt2(), with parameters $p1$ and $p2$ should be defined in such cases, with the constraint that $p1 \neq p2$.\\

Surveys involving any sensitive attribute can be conducted in this way. The survey statistician can modify the statements according to the sensitive attribute under study.\\

Similarly, any RRT can be implemented in this way. As a simple example of modification of the function, the following code is provided, which displays the colors of a rainbow depending on the outcome of a randomization device. 

\begin{verbatim}
randrt<-function()
{
r<-sample(1:7,1,replace=F,prob=c(0.1,0.2,0.1,0.3,0.1,0.1,0.1))
if (r==1) txt<-"Violet." 
else if (r==2) txt<-"Indigo." 
else if (r==3) txt<-"Blue." 
else if (r==4) txt<-"Green." 
else if (r==5) txt<-"Yellow." 
else if (r==6) txt<-"Orange." 
else if (r==7) txt<-"Red." 
return(txt)
}
\end{verbatim}

\section{Data Collection:}

This section focuses on data collection, a crucial part of surveys. The two main parts of the Shiny app are discussed, namely, the User Interface and the Server Function.

\subsection{The User Interface (UI) Object:}

The layout and appearance of the application is controlled by the User Interface (UI) Object. This can be set according to the personal preference of the survey statisticians and the requirements of the survey. A fully functional UI must consist of the following:

\begin{enumerate}
	\item Title of the survey, rendered using \begin{verbatim}
	titlePanel("Title")
	\end{verbatim}
	\item Instructions for the respondents, displayed using \begin{verbatim}
	helpText("Text")
	\end{verbatim}
	
	For example, the following instructions may be displayed to the respondents:
	
	"If the outcome of the device matches the attribute that you possess, 
	(for example: the device says *I am wearing black clothes* 
	and you are indeed wearing black clothes),
	then select Yes, else select No."
	\item Privacy statements or disclaimers, displayed using \begin{verbatim}
	helpText("Text")
	\end{verbatim}
	
	A sample disclaimer is given below:
	
	"Your response is completely anonymous. Only your Yes/No response is stored."
	\item The outcome of the randomization device(s), rendered using \begin{verbatim}
	textOutput("text")
	\end{verbatim}
	\item Interface for the respondent to answer Yes or No. The code used for this purpose is:
	
	\begin{verbatim}
	textOutput("text"), #Shows outcome of Randomization Device 
	radioButtons("resp","Do you possess the attribute that the Randomization Device says you do?",
	c("Yes"="y","No"="n")), #Takes input
	actionButton("submit","Submit",class="btn-success")
	\end{verbatim}
	In surveys involving two randomization devices, the code given above is suitably modified to display two outcomes ("text1" and "text2") and take two user inputs ("resp1" and "resp2") for the two randomization devices respectively.\\
	\item Table showing the responses received so far, displayed using \begin{verbatim}
	dataTableOutput("mydata", width = 300)
	\end{verbatim}
	
	Displaying the collected data to the respondents is optional. However, it helps to communicate to the respondent that no personal or identifying information about them is being stored. 
	\item An option of downloading the responses received so far can be given to the public using the code \begin{verbatim}
	downloadButton("download", "Download")
	\end{verbatim}
	
\end{enumerate}

\subsection{The Server function:}

The Server function contains the instructions needed to carry out the functionalities of the application. It is used for the following purposes:

\begin{enumerate}
	\item The text output of Randomization device are generated from function "randrt()" as follows:
	
	\begin{verbatim}
	output$text<-renderPrint(randrt())
	\end{verbatim}
	
	For surveys involving two randomization devices, these are suitably modified to generate two text outputs "text1" and "text2" corresponding to the two functions "randrt1()" and "rantrt2()" respectively. 	
	
	\item The following code is used to save the data in the CSV file when the user clicks the "Submit" button:
	
	\begin{verbatim}
	formData <- reactive({
	data <- sapply(fields, function(x) input[[x]])
	data
	})
	
	observeEvent(input$submit, {
	saveData(formData())
	write.csv(mydata,"data.csv",row.names = FALSE)
	})
	\end{verbatim}
	
	\item The following code is used for displaying the saved data as a table when the user clicks the "Submit" button:
	
	\begin{verbatim}
	output$mydata <- renderDataTable({
	input$submit
	loadData()
	})
	\end{verbatim}
	
	\item The option for downloading the collected data as a CSV file is provided by the following code:
	
	\begin{verbatim}
	output$download <- downloadHandler(
	filename = function() {
	paste("mydata-", Sys.Date(), ".csv", sep="")
	},
	content = function(file) {
	write.csv(mydata, file,row.names = FALSE)
	})
	\end{verbatim}
	
\end{enumerate}

\subsection{A call to the shinyApp function:}

The following code is used to create the Shiny application from the UI/server pair: 
\begin{verbatim}
shinyApp(ui = ui, server = server)
\end{verbatim}

\section{Data storage:} 
\label{storage}

The data collected during the survey can be stored locally or in a remote storage space. The advantage of storing data locally lies merely in its simplicity. However, data may be lost when the application becomes inactive or "sleeping" in the server, depending on the options provided by the server. The simplest way to overcome this is to set a pre-defined time window for the respondents to take part in the survey, and download the collected data as a CSV file at the end of such time window. This is suitable only for small scale surveys.\\

For surveys that involve a vast amount of data and involve a longer period of data collection, which is the case for almost all surveys, it is more suitable to use a reliable Remote Storage option such as Google Drive, DropBox, etc. R packages such as googlesheets4 (\cite{googlesheets}), rdrop2 (\cite{rdrop}), etc. are available for the purpose. Caution should be exercised in such cases, as it requires non-interactive authentication for the application to access the Remote Storage. \\

Applications built utilizing both local and remote storage have been discussed in details in the following sections. It is recommended that the survey statisticians further explore the storage options and pick the one that is most suitable for their needs. (\cite{storagearticle})\\

\subsection{Local storage:}
\label{local}

A single-file Shiny Web Application is created in RStudio. In the folder that contains the application, a CSV file named "data.csv" is also saved. For RRT with a single randomization device, the first row of this file contains the row heading "resp". For RRT with two randomization devices, the row headings are "resp1" and "resp2". The headings correspond to the names of the variables of interest. The data collected, i.e. the Yes/No responses will be stored in this file. This file should also be uploaded to the server that hosts the application during deployment.\\

The functions for saving and loading data are as follows:\\

The variables to be stored are first defined for single and two randomization devices as follows: 

\begin{verbatim}
fields <- c("resp") #single randomization device
fields <- c("resp1","resp2") #two randomization devices
\end{verbatim}

The file for storing the data, namely, "data.csv", located in the same folder as the application, is loaded into a data frame named "mydata" using the following code:

\begin{verbatim}
mydata<-read.csv("data.csv")
\end{verbatim}

The function for saving the collected data into this data frame is defined as follows:

\begin{verbatim}
saveData <- function(data) 
{
data <- as.data.frame(t(data))
if (exists("mydata")) 
mydata <<- rbind(mydata, data)
else
mydata <<- mydata
}
\end{verbatim}

The following function is used to load and display the data saved in the data frame created above:

\begin{verbatim}
loadData <- function() 
{
if (exists("mydata")) 
mydata
}
\end{verbatim}

Complete code for Warner's RRT (\cite{warner}) using local storage has been provided as Appendix 1.

\subsection{Data storage in Google Sheets:}
\label{gdrive}

The R package "googlesheets4" \cite{googlesheets} can be used to store the responses within a Google Sheet. The survey statisticians first have to create a Google Sheet where the responses will be stored and obtain its key or id. Data collected will be saved in this sheet using the function \begin{verbatim}
sheet_append()
\end{verbatim}

The survey statisticians also need to obtain a "token" that allows the Shiny Application to access the Google Sheet. This is done using the following code when the application is run for the first time:

\begin{verbatim}
gs4_auth(cache = ".secrets") 
\end{verbatim}

It opens up a browser window, where credentials are to be entered and access granted. The token is saved in a folder named ".secrets", which is to be uploaded to the server alongside the application when the application is deployed. Once a token has been obtained, this token is used in the subsequent sessions for non-interactive authentication by substituting the code given above with the following code: 

\begin{verbatim}
gs4_auth(cache = ".secrets", email = TRUE, use_oob = TRUE)
\end{verbatim}

Detailed instructions for non-interactive authentication is available on the web, such as (\cite{nonintauth}).\\

The code for storing data in Google Sheet from a survey involving Simmons' (1967) RRT (\cite{simmons}) with single randomization device has been provided as Appendix 2.\\

\section{Discussions:}
\label{discussions}

Although Warner (1965) (\cite{warner}) and Simmons (1965) (\cite{simmons}) RRTs have been used here as illustrations, any Randomized Response Technique can be implemented as a web-based survey using the code given in the manuscript, with the mere modification of the function "randrt()".\\

Hence, the Shiny application introduced above has wide-scale applications for survey statisticians and practitioners conducting surveys involving sensitive attributes. \\

For the purpose of demonstrating the effectiveness of the code presented in this paper, the authors have designed an electronic survey, where the data collected is stored securely in their Google Drive. The survey can be accessed at https://samplesurveyscholar.shinyapps.io/simmonsforpaper/.\\

\section*{Acknowledgment/ Funding:}

We are thankful to the Department of Science and Technology, Science \& Engineering Research Board (DST-SERB) for providing financial assistance under Grant EMR/2017/000882 and also to IIT (ISM) Dhanbad for providing the infrastructure facilities.\\

\section{Appendix 1: Warner (1965) RRT with local data storage}
\begin{verbatim}
library(shiny)

#Parameter setting for the survey statistician
#Set the probability of the sensitive attribute in the randomization device
p=0.4

mydata<-read.csv("data.csv")

fields <- c("resp")

randrt<-function()
{
r<-sample(1:2,1,replace=F,prob=c(p,1-p))
if (r==1)
txt<-"I am a smoker." #Set your sensitive attribute here
else if (r==2)
txt<-"I am not a smoker." #Set your non-sensitive attribute here
return(txt)
}


saveData <- function(data) {
data <- as.data.frame(t(data))
if (exists("mydata")) {
mydata <<- rbind(mydata, data)
} else {
mydata <<- mydata
}
}

loadData <- function() {
if (exists("mydata")) {
mydata
}
}

ui <- fluidPage(

titlePanel("Warner (1965) RRT to anonymously determine the proportion of individuals with high fever in a given population"),

sidebarLayout(
sidebarPanel(
helpText("INSTRUCTIONS:"),
helpText("PRIVACY:"), 
helpText("The Randomization Device has generated the outcome:"),
textOutput("text"),

radioButtons("resp","Do you possess the attribute that the Randomization Device says you do?",
c("Yes"="y","No"="n")),
actionButton("submit","Submit",class="btn-success"),
width=6
),
mainPanel(
dataTableOutput("mydata", width = 300), tags$hr(),
downloadButton("download", "Download"),
width=4
)
)
)


server <- function(input, output, session) {

output$text<-renderPrint(randrt())


formData <- reactive({
data <- sapply(fields, function(x) input[[x]])
data
})

observeEvent(input$submit, {
saveData(formData())
write.csv(mydata,"data.csv",row.names = FALSE)
})

output$mydata <- renderDataTable({
input$submit
loadData()
})

#thedata <- reactive(mydata)

output$download <- downloadHandler(
filename = function() {
paste("mydata-", Sys.Date(), ".csv", sep="")
},
content = function(file) {
write.csv(mydata, file,row.names = FALSE)
})

}

# Run the application 
shinyApp(ui = ui, server = server)

\end{verbatim}

\section{Appendix 2: Simmons (1967) RRT with remote data storage}

\begin{verbatim}
library(shiny)
library(shinydashboard)
library(googlesheets4)
library(googledrive)


gs4_auth(cache = ".secrets") #for the first time running the app in R to get the OAuth token
gs4_auth(cache = ".secrets", email = TRUE, use_oob = TRUE)
fields <- c("resp")

randrt<-function()
{
r<-sample(1:2,1,replace=F,prob=c(0.4,0.6))
if (r==1)
{
txt<-"I have high fever."
}
else if (r==2)
{
txt<-"I was born on a Sunday."
}
return(txt)
}

ui <- fluidPage(

titlePanel("Simmons (1967) RRT to anonymously determine the proportion of smokers in a given population"),

sidebarLayout(
sidebarPanel(
helpText("INSTRUCTIONS"),
helpText("PRIVACY"), 
helpText("The Randomization Device has generated the outcome:"),
textOutput("text"),
width=6
),
mainPanel(
radioButtons("resp","Do you possess the attribute that the Randomization Device says you do?",
c("Yes"="y","No"="n")),
actionButton("submit","Submit",class="btn-success"),
width=4
)
)
)

server <- function(input, output, session) {

output$text<-renderPrint(randrt())
observeEvent(input$submit,{
sheet_append("XXX",as.data.frame(input$resp)) #XXX is Google Sheet key

}) 


}

# Run the application 
shinyApp(ui = ui, server = server)


\end{verbatim}


\end{document}